\input harvmac
\input epsf
\def\Title#1#2{\rightline{#1}\ifx\answ\bigans\nopagenumbers\pageno0\vskip1in
\else\pageno1\vskip.8in\fi \centerline{\titlefont #2}\vskip .5in}
%

%\voffset=.75in
%\hoffset=.7in
%%%%%%%%%%%%%%%%%%
%
% Figure macros, SBG 5/93
%
\ifx\epsfbox\UnDeFiNeD\message{(NO epsf.tex, FIGURES WILL BE IGNORED)}
\def\figin#1{\vskip2in}% blank space instead
\else\message{(FIGURES WILL BE INCLUDED)}\def\figin#1{#1}%\epsfverbosetrue
\fi
\def\Fig#1{Fig.~\the\figno\xdef#1{Fig.~\the\figno}\global\advance\figno
 by1}
%
%  ifig   usage:
%
%         \ifig\figlabel{caption}{figfile}{vsize}
%
% where vsize is the desired vertical size of the figure in truein
%
\def\ifig#1#2#3#4{
\goodbreak\midinsert
\figin{\centerline{\epsfysize=#4truein\epsfbox{#3}}}
\narrower\narrower\noindent{\footnotefont
{\bf #1:}  #2\par}
\endinsert
}

%%%%%%%%%%%%%%%%%%
%
% Figure macros, SBG 3/03
%
%\ifx\includegraphics\UnDeFiNeD\message{(NO graphicx.tex, FIGURES WILL BE IGNORED)}
%\def\figin#1{\vskip2in}% blank space instead
%\else\message{(FIGURES WILL BE INCLUDED)}\def\figin#1{#1}
%\fi
%\def\Fig#1{Fig.~\the\figno\xdef#1{Fig.~\the\figno}\global\advance\figno
% by1}
%
%  Ifig   usage:
%
%         \Ifig{\Fig\figlabel}{caption}{figfile}{hsize}
%
% where vsize is the desired vertical size of the figure in truein
%
%\def\Ifig#1#2#3#4{
%\goodbreak\midinsert
%\figin{\centerline{
%\includegraphics[width=#4truein]{#3}}}
%\narrower\narrower\noindent{\footnotefont
%{\bf #1:}  #2\par}
%\endinsert
%}
%
%defs
%
\font\ticp=cmcsc10

\def\ajou#1&#2(#3){\ \sl#1\bf#2\rm(19#3)}
\def\jou#1&#2(#3){,\ \sl#1\bf#2\rm(19#3)}

\def\frac#1#2{{#1\over#2}}

\def\cL{{\cal L}}

\def\cR{{\cal R}}
\def\cC{{\cal C}}

\overfullrule=0pt
%
%refs
%
\lref\wmap{
C.~L.~Bennett {\it et al.},
``First Year Wilkinson Microwave Anisotropy Probe (WMAP) Observations: Preliminary Maps and Basic Results,''
arXiv:astro-ph/0302207.
%%CITATION = ASTRO-PH 0302207;%%
}
\lref\Bankslittle{
T.~Banks,
``Cosmological breaking of supersymmetry or little Lambda goes back to  the future. II,''
arXiv:hep-th/0007146.
%%CITATION = HEP-TH 0007146;%%
}
\lref\Fischler{W. Fischler, unpublished (2000)\semi
W. Fischler, ``Taking de Sitter seriously," Talk given at
Role of Scaling Laws in Physics and Biology (Celebrating
the 60th Birthday of Geoffrey West), Santa Fe, Dec. 2000.}
\lref\Wittobs{
E.~Witten,
``Quantum gravity in de Sitter space,''
arXiv:hep-th/0106109.
%%CITATION = HEP-TH 0106109;%%
}
\lref\Astrods{
A.~Strominger,
``The dS/CFT correspondence,''
JHEP {\bf 0110}, 034 (2001)
[arXiv:hep-th/0106113]
%%CITATION = HEP-TH 0106113;%%
\semi
M.~Spradlin, A.~Strominger and A.~Volovich,
``Les Houches lectures on de Sitter space,''
arXiv:hep-th/0110007.
%%CITATION = HEP-TH 0110007;%%
}
\lref\Sussrecur{
L.~Dyson, J.~Lindesay and L.~Susskind,
``Is there really a de Sitter/CFT duality,''
JHEP {\bf 0208}, 045 (2002)
[arXiv:hep-th/0202163].
%%CITATION = HEP-TH 0202163;%%
}
\lref\DysonPF{
L.~Dyson, M.~Kleban and L.~Susskind,
``Disturbing implications of a cosmological constant,''
JHEP {\bf 0210}, 011 (2002)
[arXiv:hep-th/0208013].
%%CITATION = HEP-TH 0208013;%%
}
\lref\GKS{
N.~Goheer, M.~Kleban and L.~Susskind,
``The trouble with de Sitter space,''
arXiv:hep-th/0212209.
%%CITATION = HEP-TH 0212209;%%
}
\lref\Acha{B.S. Acharya, ``A moduli fixing mechanism in M theory," hep-th/0212294.
}
\lref\HKS{
S.~Hellerman, N.~Kaloper and L.~Susskind,
``String theory and quintessence,''
JHEP {\bf 0106}, 003 (2001)
[arXiv:hep-th/0104180].
%%CITATION = HEP-TH 0104180;%%
}
\lref\FKMP{
W.~Fischler, A.~Kashani-Poor, R.~McNees and S.~Paban,
``The acceleration of the universe, a challenge for string theory,''
JHEP {\bf 0107}, 003 (2001)
[arXiv:hep-th/0104181].
%%CITATION = HEP-TH 0104181;%%
}
\lref\Sethi{
K.~Dasgupta, G.~Rajesh and S.~Sethi,
``M theory, orientifolds and G-flux,''
JHEP {\bf 9908}, 023 (1999)
[arXiv:hep-th/9908088].
%%CITATION = HEP-TH 9908088;%%
}
\lref\GVW{
S.~Gukov, C.~Vafa and E.~Witten,
``CFT's from Calabi-Yau four-folds,''
Nucl.\ Phys.\ B {\bf 584}, 69 (2000)
[Erratum-ibid.\ B {\bf 608}, 477 (2001)]
[arXiv:hep-th/9906070].
%%CITATION = HEP-TH 9906070;%%
}
\lref\GKP{
S.~B.~Giddings, S.~Kachru and J.~Polchinski,
``Hierarchies from fluxes in string compactifications,''
Phys.\ Rev.\ D {\bf 66}, 106006 (2002)
[arXiv:hep-th/0105097].
%%CITATION = HEP-TH 0105097;%%
}
\lref\BBHL{
K.~Becker, M.~Becker, M.~Haack and J.~Louis,
``Supersymmetry breaking and alpha'-corrections to flux induced  potentials,''
JHEP {\bf 0206}, 060 (2002)
[arXiv:hep-th/0204254].
%%CITATION = HEP-TH 0204254;%%
}
\lref\Silv{
E.~Silverstein,
``(A)dS backgrounds from asymmetric orientifolds,''
arXiv:hep-th/0106209.
%%CITATION = HEP-TH 0106209;%%
}
\lref\KKLT{
S.~Kachru, R.~Kallosh, A.~Linde and S.~P.~Trivedi,
``De Sitter vacua in string theory,''
arXiv:hep-th/0301240.
%%CITATION = HEP-TH 0301240;%%
}
\lref\DiSe{
M.~Dine and N.~Seiberg,
``Is The Superstring Weakly Coupled?,''
Phys.\ Lett.\ B {\bf 162}, 299 (1985).
%%CITATION = PHLTA,B162,299;%%
}
\lref\Sussanth{
L.~Susskind,
``The Anthropic Landscape of String Theory,''
arXiv:hep-th/0302219.
%%CITATION = HEP-TH 0302219;%%
}
\lref\AdStunnel{
S.~R.~Coleman and F.~De Luccia,
``Gravitational Effects On And Of Vacuum Decay,''
Phys.\ Rev.\ D {\bf 21}, 3305 (1980).
%%CITATION = PHRVA,D21,3305;%%
}
\lref\RSI{
L.~Randall and R.~Sundrum,
``A large mass hierarchy from a small extra dimension,''
Phys.\ Rev.\ Lett.\  {\bf 83}, 3370 (1999)
[arXiv:hep-ph/9905221].
%%CITATION = HEP-PH 9905221;%%
}
\lref\DeGi{
O.~DeWolfe and S.~B.~Giddings,
``Scales and hierarchies in warped compactifications and brane worlds,''
arXiv:hep-th/0208123.
%%CITATION = HEP-TH 0208123;%%
}
\lref\Banksinstab{
T.~Banks,
``Heretics of the false vacuum: Gravitational effects on and of vacuum  decay. II,''
arXiv:hep-th/0211160.
%%CITATION = HEP-TH 0211160;%%
}
\lref\GKPunpub{S.B. Giddings, S. Kachru, and J. Polchinski, unpublished.}
\lref\Wittendthree{
E.~Witten,
``Non-Perturbative Superpotentials In String Theory,''
Nucl.\ Phys.\ B {\bf 474}, 343 (1996)
[arXiv:hep-th/9604030].
%%CITATION = HEP-TH 9604030;%%
}
\lref\Casref{
P.~Candelas and S.~Weinberg,
``Calculation Of Gauge Couplings And Compact Circumferences From Selfconsistent Dimensional Reduction,''
Nucl.\ Phys.\ B {\bf 237}, 397 (1984).
%%CITATION = NUPHA,B237,397;%%
}
\lref\RaPe{
B.~Ratra and P.~J.~Peebles,
``Cosmological Consequences Of A Rolling Homogeneous Scalar Field,''
Phys.\ Rev.\ D {\bf 37}, 3406 (1988).
%%CITATION = PHRVA,D37,3406;%%
}
\lref\BoPo{
R.~Bousso and J.~Polchinski,
``Quantization of four-form fluxes and dynamical neutralization of the  cosmological constant,''
JHEP {\bf 0006}, 006 (2000)
[arXiv:hep-th/0004134].
%%CITATION = HEP-TH 0004134;%%
}
\lref\Banksobs{
T.~Banks and W.~Fischler,
``M-theory observables for cosmological space-times,''
arXiv:hep-th/0102077\semi
%%CITATION = HEP-TH 0102077;%%
T.~Banks, W.~Fischler and S.~Paban,
``Recurrent nightmares?: Measurement theory in de Sitter space,''
JHEP {\bf 0212}, 062 (2002)
[arXiv:hep-th/0210160].
%%CITATION = HEP-TH 0210160;%%
}
\lref\KaVe{
S.~Kachru, J.~Pearson and H.~Verlinde,
``Brane/flux annihilation and the string dual of a non-supersymmetric  field theory,''
JHEP {\bf 0206}, 021 (2002)
[arXiv:hep-th/0112197].
%%CITATION = HEP-TH 0112197;%%
}
\lref\CHT{S.M. Carroll, M. Hoffman, and M. Trodden, ``Can the dark energy equation-of-state parameter $w$ be less than -1?'' [arXiv:astro-ph/0301273].}
\lref\Star{
A. A. Starobinsky, ``Stochastic De Sitter (Inflationary)
Stage In The Early Universe," in: {\sl Current Topics in
Field Theory, Quantum Gravity and Strings,} Lecture
Notes in Physics, eds. H.J. de Vega and N. Sanchez
(Springer, Heidelberg 1986) 206, p. 107.}

\Title{\vbox{\baselineskip12pt
\hbox{hep-th/0303031}\hbox{MIFP-03-03}
}}
{\vbox{\centerline{
The fate of four dimensions}
}}
\centerline{{\ticp Steven B. Giddings}\footnote{$^\dagger$}
{Email address:
giddings@physics.ucsb.edu} }
\bigskip\centerline{ {\sl Department of Physics}\footnote{$^*$}{Primary address.}}
\centerline{\sl University of California}
\centerline{\sl Santa Barbara, CA 93106-9530}
\centerline{and}
\centerline{\sl George P. \& Cynthia W. Mitchell Institute for Fundamental Physics}
\centerline{\sl Texas A\&M University}
\centerline{\sl College Station, TX 77843-4242}
\bigskip\bigskip
\centerline{\bf Abstract}

In gravitational theories with extra dimensions, it is argued that the existence of a positive vacuum energy generically implies catastrophic instability of our four-dimensional world.  The most generic instability is a decompactification transition to growth of the extra dimensions, although other equally bad transitions may take place.  This follows from simple considerations based on the form of the potential for the size modulus of the extra dimensions, and apparently offers a resolution of the conundrums presented by eternal de Sitter space.  This argument is  illustrated in the context of string theory with a general discussion of potentials generated by fluxes, wrapped branes, and stringy corrections.  Moreover, it is unlikely that the present acceleration of the Universe represents an ongoing transition in a  quintessence scenario rolling towards decompactification, unless the higher dimensional theory has a cosmological constant.

%\draftmode
\Date{}

\newsec{Introduction}

The reality of a small positive vacuum energy is becoming increasingly harder to deny; the recent WMAP results\refs{\wmap} lend even more solidity to the statement that the dominant energy density in the Universe is either the cosmological constant, or a similar negative pressure component. From the theoretical viewpoint, this presents a serious challenge.  The small observed value sharpens the challenge considerably.  Theory typically predicts either a vanishing cosmological constant, but at the same time unrealistic phenomenology such as unbroken supersymmetry, or, with broken supersymmetry, predicts a cosmological constant differing by a factor at least $10^{60}$ from that apparently implied by observation.

Moreover, it has been argued that a nonzero cosmological constant presents even more profound challenges of principle in any consistent theory of quantum gravity.  The work of recent years has strongly suggested that we take the Bekenstein-Hawking entropy of a black hole as a literal count of the number its microstates.  De Sitter space likewise has a finite entropy, and this has been argued to imply that any microphysical description of it will have only finitely many states\refs{\Bankslittle,\Fischler}.  Such a theory would be very different from quantum field theory or string theory.

Furthermore, the prospect of living in eternal de Sitter space raises a number of challenges to a theory of observeables\refs{\Wittobs,\Astrods,\Banksobs}, to theories of initial conditions, and even to consistency\refs{\Sussrecur,\DysonPF,\GKS}.  In particular, \Sussrecur\ has argued that, given a finite number of states for dS space, physics must exhibit Poincar\'e recurrences on a timescale 
\eqn\Poinc{T_{recur} \sim e^{S_{dS}}}
where $S_{dS}$ is the de Sitter entropy.  Furthermore, ref.~\GKS\ argues that there is no obvious way to realize a system with a finite entropy and the symmetries of dS space.  In short, an eternally inflating Universe presents a number of thorny problems.

These observations can be coupled with a pragmatic one:  in today's dominant paradigm for a theory of quantum gravity, string theory, it has been very difficult to find de Sitter vacua, or even models for quintessence\refs{\HKS,\FKMP}.  (But for a construction in non-critical string theory, see \refs{\Silv}.)

At the same time, gravity theories with extra dimensions generally exhibit a related set of problems.  Compactifications of extra dimensions typically have moduli parameters which parametrize the size and shape of the extra dimensions.   It is in general difficult to find physics that convincingly fixes these moduli to reasonable values.  This has in particular been a long-standing problem in string theory.  However, significant progress has been made on this problem in the past few years, with the realization that trapped fluxes present in the extra dimensions can generate a potential for these moduli\refs{\Sethi,\GVW,\GKP}.  In particular, ref.~\GKP\ constructed classical solutions of string theory where such fluxes can generate a potential for all but one of the moduli, the overall scale of the internal manifold, or radial dilaton.  It was realized that higher-order string corrections would generate a potential for the radial dilaton\GKP, and some of these corrections were first explicitly found in ref.~\refs{\BBHL}.  

In the context of a theory of extra dimensions that generates a potential for the moduli, and in particular for the radial dilaton, the vacuum energy at a minimum of this potential corresponds to the cosmological constant.  This means that the problem of fixing these moduli and that of determining the value of the cosmological constant are linked.  The question becomes one of understanding whether the potential for moduli has a minimum that generates the correct cosmological constant.  But there is a tension here -- if one {\it were} to find a stable minimum giving rise to dS space, this would seemingly encounter immediate conflict with the conundrums presented by the finite entropy of dS space.  

Recently further progress on the moduli-fixing problem has been made by Kachru, Kallosh, Linde, and Trivedi\refs{\KKLT}, who added anti-D3 branes to the flux compactifications of \refs{\GKP}, breaking supersymmetry.  This, together with accounting for stringy corrections to the potential of \GKP, produces a potential with all moduli fixed at a de Sitter minimum.  However, this minimum is only {\it metastable}:  it is unstable to either quantum tunneling or thermal excitation over a barrier, after which time the Universe runs away to infinity in moduli space.  Moreover, the authors of \KKLT\ argued that the instability operates on a time scale that will be short as compared to the Poincar\'e time scale.
 
There has been a general awareness, based on extrapolation of an old argument of Dine and Seiberg\refs{\DiSe},\foot{Dine and Seiberg's original argument was made for the string dilaton.}
that any potential that is generated for moduli in string theory
will have runaway directions to infinite flat supersymmetric ten-dimensional space.  The reason is that there is apparently no mechanism in string theory that could prevent ten-dimensional space from being an exact solution to the theory.

Goheer, Kleban, and Susskind have reiterated these arguments\refs{\GKS}, and in particular describe a simple argument\refs{\Star,\KKLT} that the lifetime of such a metastable de Sitter minimum is always less than the Poincar\'e recurrence time.\foot{These arguments have been amplified in \refs{\Sussanth}, received while this paper was in preparation.}

This apparently circumvents the problems presented by recurrences, and by the finite entropy of de Sitter space.  In such a framework, where de Sitter space is not eternal, 
de Sitter space may correspond to a finite number of states, but these are among the infinite number of states of infinite flat space. De Sitter space is merely a metastable excitation of the infinite number of degrees of freedom of this space.  The width of the de Sitter states makes it impossible to find any physical consequences of the smaller spacing between the finite number of states of the de Sitter space.

This paper will go on to provide a broader context for these observations, based on a semiclassical analysis, to compactifications of a general gravitational theory of extra dimensions.  Specifically, it will give a very general argument for a problem with finding gravitational models with extra dimensions and stable de Sitter minima.  The essential point is that for a generic compactification, the potential for the radial dilaton must always vanish at infinity.  This means that a de Sitter minimum must always be unstable, generically either to runaway to infinitely expanded extra dimensions, or, if there is an intervening negative energy minimum in the potential, to gravitational collapse to a big crunch. (A third possibility, in cases where there are also exactly supersymmetric four-dimensional vacua, will be discussed in the following.)  Moreover, it appears likely that such an argument should extend to forbid stable four-dimensional de Sitter vacuua even in a region where a semiclassical analysis of the compact theory is not possible.

In short, the analysis of this paper suggests that if 1) there are extra dimensions of space and 2) the Universe is undergoing accelerated expansion, then the present four-dimensional state of the Universe is not a stable state.  The Universe is catastrophically unstable either to decompactification of extra dimensions, to gravitational collapse to a big crunch, or in special cases, possibly to decay to a four-dimensional supersymmetric universe.

The next section will first study the general form of the radial dilaton potential, and give a general argument for this instability.   This will be followed by a description of some examples of radion potentials, induced by fluxes threading extra dimensions, wrapped branes, and string-coupling and sigma model corrections.  Next, the possibility of deriving quintessence models is revisited and it is argued that the only known mechanism for generating a quintessential potential for the radial dilaton is the presence of a higher-dimensional cosmological constant.  The closing section contains discussion of the consequences of these arguments, together with conclusions.

\newsec{Instability of extra dimensions}

In this section I present a general argument that, if the Universe is currently in a de Sitter phase, and there are geometrical extra dimensions of spacetime, then the present de Sitter phase is a metastable state, and has a general instability towards a true ground state corresponding either to infinitely expanded extra dimensions, to an anti-de Sitter vacuum, or possibly to a supersymmetric four-dimensional theory.  In the second case, the Universe cannot reach this state, but instead ultimately undergoes uncontrolled gravitational collapse\refs{\AdStunnel}; in the latter case the instability would be equally catastrophic to physics as we know it.  Of course, in a particular theory there may equally well be less generic instabilities towards other configurations.

The assumption that the extra dimensions have a geometrical description is tantamount to assuming that the problem can be treated in the supergravity approximation.  If the supergravity approximation were {\it not} valid, then this would be an indicator that description of the extra dimensions as a semiclassical geometry had failed.  
Therefore, we consider a lagrangian of the form
\eqn\Act{S=\int d^DX \sqrt{-G}\left[ M_D^{D-2} \cR + \cL(\psi) + {\hat \cL}(\cR)\right]\ .}
Here $X$ and $G$ are the coordinates and metric of the full $D=4+d$ dimensional spacetime, $M_D$ is the $D$-dimensional Planck mass, $\cR$ is the Ricci scalar, and $ \cL(\psi)$ is the lagrangian representing the contribution of generic matter sources,
possibly including  localized sources such as D-branes.  Finally, ${\hat \cL}(\cR)$ summarizes possible corrections to the Einstein-Hilbert lagrangian that involve higher powers of the curvature.  (There also could be terms mixing $\psi$ and $\cR$, but we don't explicitly display them without jeopardizing the underlying argument.)

Suppose that there exists a solution of this lagrangian which gives de Sitter space.  This will take the form:
\eqn\metric{ds^2 = e^{2A(y)} ds_4^2 + g_{mn}(y) dy^m dy^n \ ,}
\eqn\psisol{ \psi = \psi_0(y)}
where $x$ are the 4d coordinates and $y$ are the compact coordinates, and, to preserve the de Sitter symmetry of the vacuum, the matter fields are independent of 4d coordinates.  In general we must allow for the presence of a warp factor, $A(y)$, which can play an important role in deriving phenomenologically relevant scales\refs{\RSI,\GKP,\DeGi}. 

Now consider a different set of configurations, where the compact metric is rescaled by a radial dilaton:
\eqn\rescale{ds^2 = e^{2A(y)} ds_4^2 + R^2(x)g_{mn}(y) dy^m dy^n \ .}
Of course, the function $\psi_0$ no longer satisfies the equations of the matter fields.  But, if we assume that $R(x)=e^{D(x)}$ is slowly varying on scales of order the compactification size, the matter equations can be solved for a different solution, $\psi_D$. We will discuss simple examples shortly.  Through the action \Act, this then contributes to an effective potential for the radial dilaton.

For simplicity we examine this dynamics in the context of a trivial warp factor, $A=0$; the results are readily generalized to nontrivial $A$.  This dynamics is governed by reduction of \Act.  The Einstein-Hilbert term gives
\eqn\ehred{S_{EH}= M_D^{D-2} V_d \int d^4x \sqrt{-g_4} \left[ e^{dD(x)} \cR_4 + d(d-1) (\nabla D)^2 e^{dD(x)} + e^{(d-2)D} \cR_d\right]\ ,}
where $V_d$ and $\cR_d$ are the volume and curvature of the $d$-dimensional compact metric $g_{mn}$.  In these units, the effective four-dimensional Planck mass varies with $D$.  We choose new units by the Weyl rescaling
\eqn\weylr{g_{4\mu\nu} \rightarrow e^{-dD} g_{4\mu\nu}\ }
to eliminate this dependence, and define the four-dimensional Planck mass
\eqn\mfour{M_4^2 = M_D^{D-2} V_d\ .}
The action \Act\ then becomes
\eqn\redact{\eqalign{ S = \int d^4 x \sqrt{-g_4} \Biggl\{ & M_4^2\cR_4 - {1\over 2} M_4^2 d(d+2)(\nabla D)^2 \cr &+ e^{-dD} \int d^d y \sqrt{g_d} \left[\cL(e^{2D} g_d, 
\psi_D) + M_D^{D-2}e^{-2D} \cR_d + {\hat \cL} \right] \Biggr\}\ .}}
The last set of terms provide an effective potential $V(D)$ for the radial dilation (together with possible higher-derivative terms from ${\hat \cL}$).  

We now come to a key point.  The effective potential for $D$ contains a factor that {\it falls} like the {\it inverse volume} of the compact manifold.  The only way for the potential to approach a constant, or grow, at large volume is if the lagrangian of the other fields {\it grows} at least as fast as the volume at large volume.  

The internal action corresponds to the energy of the field configuration on the compact space, at a fixed size $R$.  Therefore a constant or growing potential requires an energy density that grows as the volume.  There are no known examples of physical phenomena with such strong growth in field theory or in string theory.  Intuitively, any non-trivial field configuration will only dilute, and thus have decreasing energy density, as we expand the volume it is confined within.  As it dilutes it leaves behind vacuum, but vacuum energy density is expected to become at most
constant at large distances, or decline from decreasing numbers of degrees of freedom, in correspondence with c-theorem intuition.  Even a wrapped brane has at most an asymptotically constant energy density.  We can also think of this from the perspective of the equation of state of the field configuration on the compact space.  An energy density that grows faster than the volume corresponds to an equation of state parameter in $p=w\rho$ that is smaller than $w=-2$. 
This violates the null dominant energy condition, and there is no known physics that can give such values without introducing acausality or other instabilities.\foot{For some recent discussion, see \refs{\CHT}.}
We will investigate simple examples of physical asymptotic behavior in the next section.  

\ifig{\Fig\figone}{A generic potential with a dS minimum has a runaway direction to infinite volume}{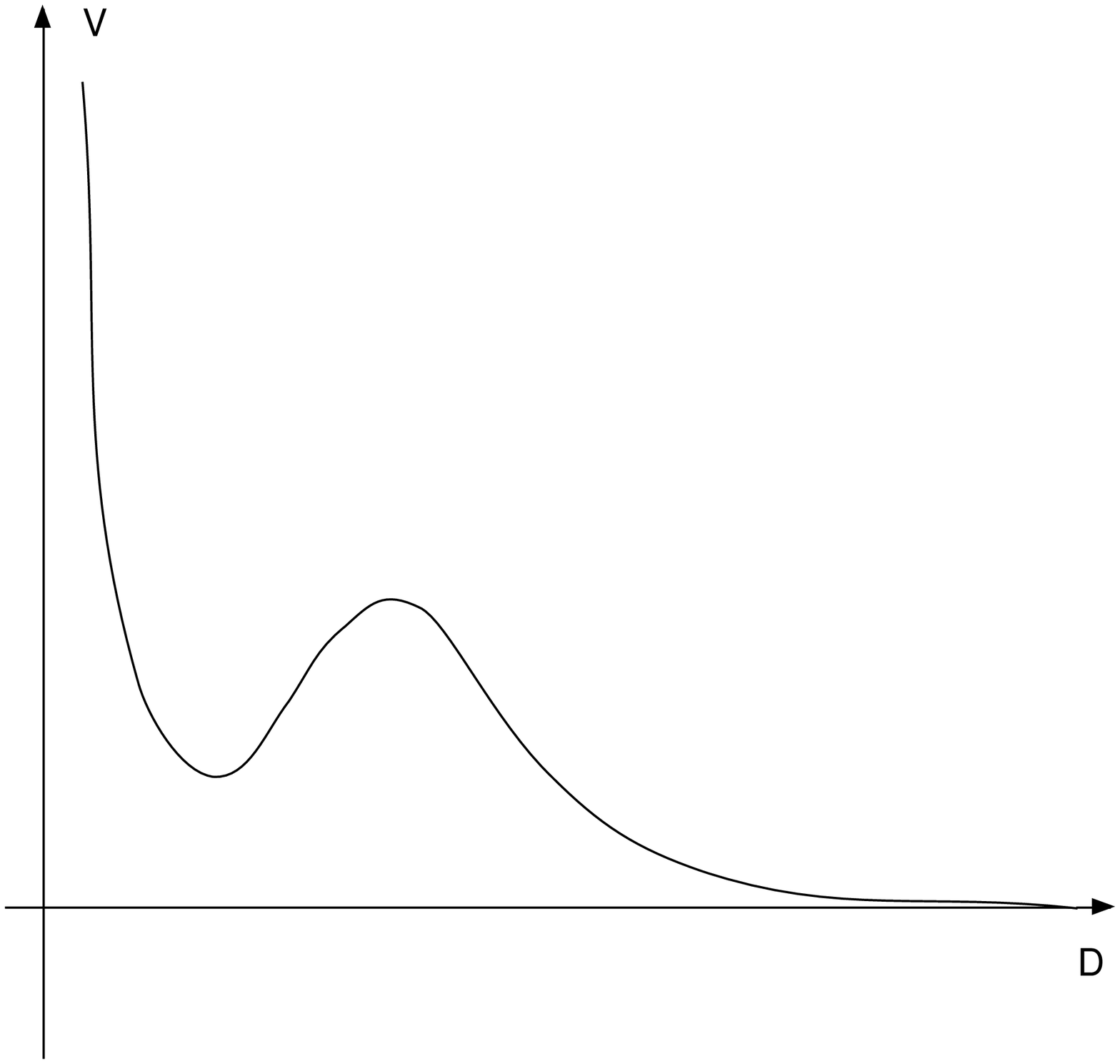}{3.5}

Thus the potential for the size of the compact dimensions must {\it generically vanish} at large volume.  If the present expansion of the Universe represents a de Sitter phase, and there are extra dimensions, we generically expect our de Sitter vacuum to be unstable to decay via tunneling.

\ifig{\Fig\figtwo}{Potentials may have a cascade of dS minima before reaching infinite volume.}{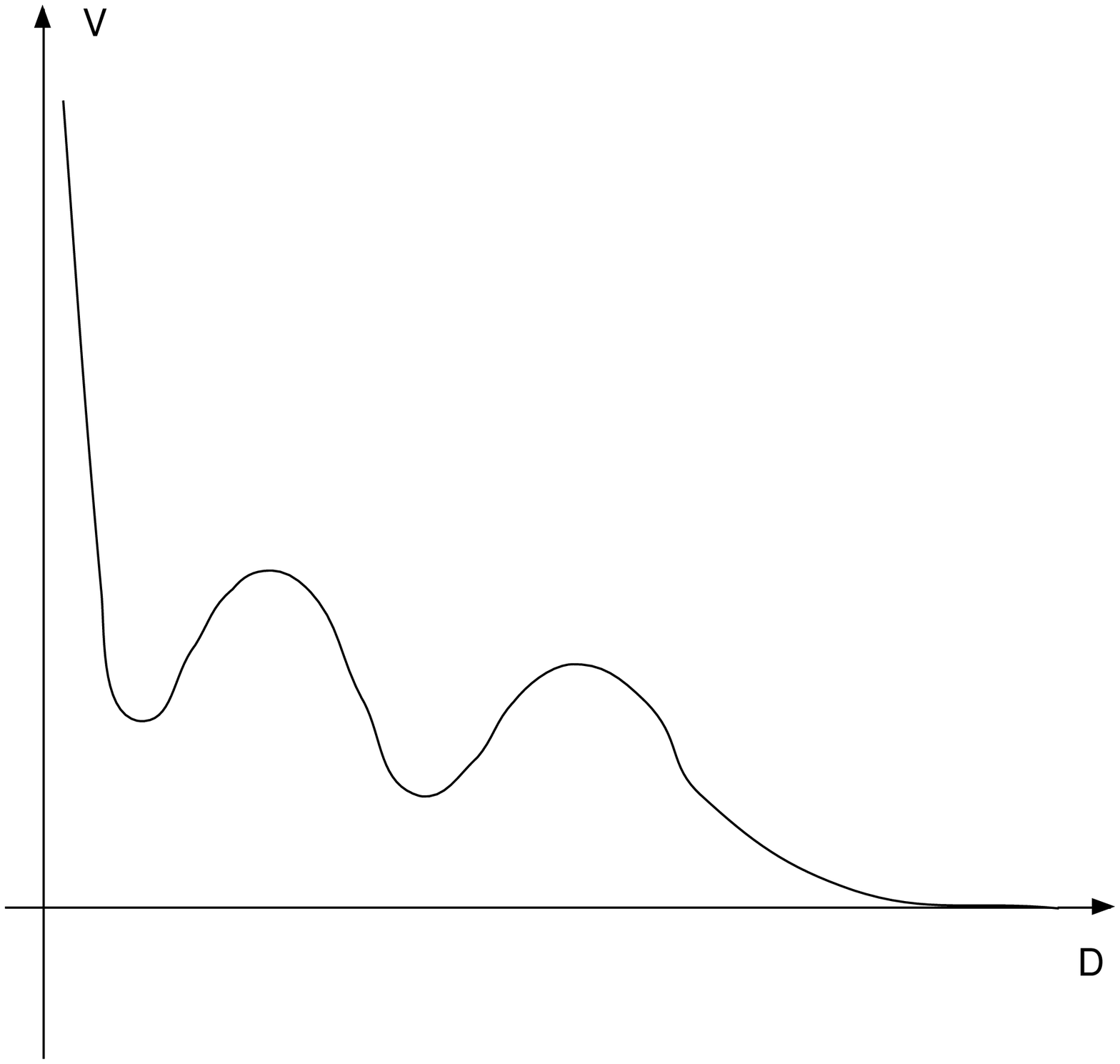}{3.5}

\ifig{\Fig\figthree}{A third alternative is an AdS minimum intervening before infinite volume.}{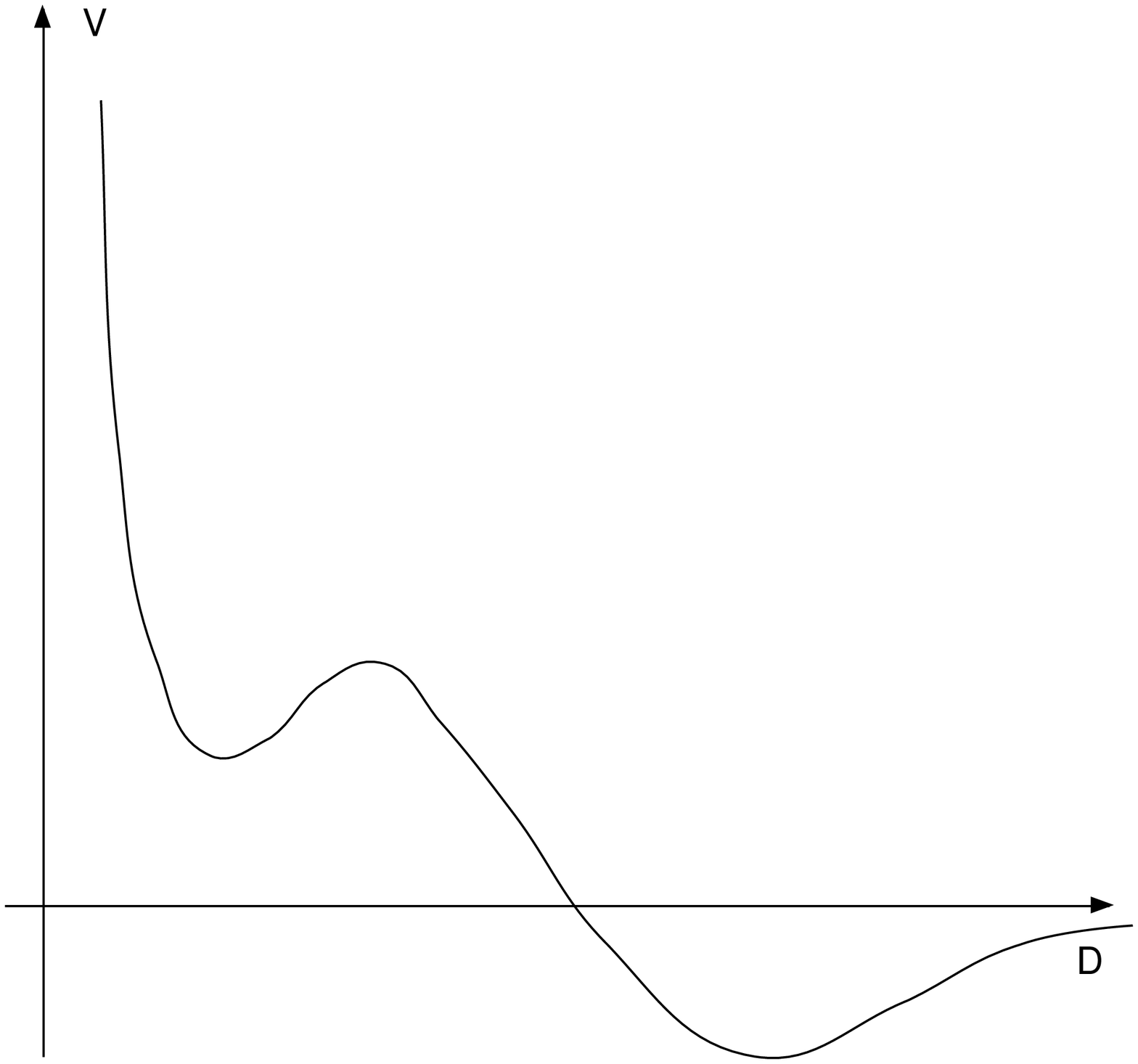}{3.5}

While we always expect the potential for the dilaton to vanish at infinity, there are a wide variety of possibilities for how it does so, for example given in figures 1-3.  If there is a de Sitter minimum, the most generic case is a potential as shown in fig.~1, where our Universe would tunnel or thermally fluctuate  into a phase in which the extra dimensions of space decompactify.  With additional terms in the potential, one could also have intermediate tunneling events into de Sitter Universes with smaller cosmological constant, as in figure 2, or if the potential takes the form of figure 3, the instability is generically to tunneling into an anti-de Sitter minimum.  As has been recently emphasized in \refs{\Banksinstab}, the latter case indeed represents a serious instability, as it will lead to a collapsing Robertson-Walker cosmology with a cosmological singularity.  The bottom line is that if there are extra dimensions and our present cosmology is de Sitter, then our four-dimensional Universe is generically ultimately unstable.

\newsec{Examples of radion potentials}

We next illustrate these general comments by cataloging some simple examples of 
possible behavior for radion potentials.  A wide class of examples are provided by string theory.  There are several sources of radion potentials in string theory.  At the classical level, such potentials are induced by fluxes or wrapped branes.  Moreover, quantum corrections in the sigma-model or string coupling expansion also generically contribute to the potential.  We consider these in turn.

\subsec{Fluxes}

As a first example from string/M theory, consider the case where we turn on some of the flux fields in the extra dimensions.  This is a general mechanism for fixing moduli\refs{\Sethi,\GVW}, as was illustrated in flux compactifications of the IIB string in \refs{\GKP}.  String theory has $p$-form field strengths, both Ramond-Ramond and Neveu-Schwarz; we consider the general case of a $p$-form generically denoted by $F_p = dA_{p-1}$.  These have kinetic actions of the form
\eqn\formact{S_p = -{1\over 4\kappa_{10}^2}\int d^{d+4} X \sqrt{-G} {F_p^2\over p!}\ ,}
perhaps with some string dilaton prefactor.  In order to preserve maximal symmetry of four dimensions, only the following components of the fluxes are allowed to be nonvanishing:
\eqn\fluxint{F_{m_1\cdots m_p}\quad\quad\quad\quad\ \  p<d}
or
\eqn\fluxext{F_{\mu\nu\lambda\sigma m_1\cdots m_{p-4}}\quad\quad p>4\ ,}
the latter being proportional to the four-dimensional volume form.  In the latter case, we can work in terms of the dual field.  Thus we only consider fluxes of the form \fluxint.

The fluxes in general obey quantization conditions.  These take the form
\eqn\quantcond{\int_{\cC_p} F_p = \mu_{8-p}\ ,}
where ${\cC_p} $ is a $p$ cycle in the spacetime manifold and $\mu_{8-p}$ is the quantized D brane charge.  This implies that $F_p$ does not scale with the radial dilaton $R$.  The dependence of the action on the radial dilaton is thus
\eqn\fluxact{S_F \propto R^{d-2p}\ .}
Combining this with \redact\ gives a radial dilation potential of the form
\eqn\fluxpot{V_F\propto R^{-d-2p} }
from a $p$ form with components in the compact directions.  This falls with increasing volume.

\subsec{Wrapped branes}

A second origin for radial dilaton potentials is to have some number of branes (or orientifold planes) wrapping the extra dimensions.  In order not to spoil the maximal four-dimensional symmetry, these branes should be extended over the non-compact four-dimensional manifold $M_4$.

At leading order in large volume, the action due to such a p-brane is just the induced Dirac-Born-Infeld action, which gives the brane's world-volume:
\eqn\braneact{S_p =-{\mu_p\over g_s} \int_{M_4\times \cC } dV_{p+1}\ ,}
where $\cC$ is the cycle over which the brane wraps, and $dV_{p+1}$ is the infinitesimal world-volume element.  

The dependence of the action on the radial dilation then immediately follows; we find
\eqn\branedil{S_p\propto \int d^4 x \sqrt{-g_4} R^{p-3} V_\cC\ .}
After the rescaling \weylr, this gives a potential
\eqn\branepot{ V_p \propto R^{p-3-2d } }
for the radial dilaton.  This potential may take either sign:  positive for a D brane, but negative with negative tension sources such as orientifold planes -- this can help to generate non-trivial minima\refs{\GKPunpub,\Silv}.\foot{For a related apporach to moduli fixing see \refs{\Acha}.}  It too falls to zero for increasing $R$; the least rapid decline is from a fully space-filling brane, $p=3+d$, and gives a potential that falls with the inverse volume of the compact dimensions.  

\subsec{String corrections}

In string theory,  corrections to a given configuration arise both in the $\alpha'$ expansion and in the string-loop expansion.  Both kinds of corrections have been previously investigated in the literature.  

\noindent{3.3.1 Loop corrections}

For example, \refs{\KKLT} made use of non-perturbative corrections in the string loop expansion.  
These are induced by euclidean D3 brane instantons\refs{\Wittendthree}, or by gluino condensation on stacks of D7 branes\KKLT.  
These correct the 4d effective superpotential that arises in the flux compactifications of \GKP.  The resulting superpotential takes the form 
\eqn\nonpertW{W= Ae^{-aR^4}\ ,}
which vanishes exponentially rapidly at large volume.

More generally, in a typical non-supersymmetric compactification of a higher-dimensional theory, one may have perturbative corrections in the coupling constants of the underlying theory.  The generic leading correction of this form is the one loop correction giving the Casimir energy.  Casimir effects have been advocated as a means of stabilizing extra dimensions since the early days of Kaluza-Klein theories\refs{\Casref}.  However, Casimir energies always fall like a power of the radius of the extra dimensions, again leading to a potential that vanishes asymptotically.

\noindent{3.3.2 String corrections}

String theory also leads to {\it classical} corrections to the action, at increasing orders in the $\alpha'$ expansion.  These are generically terms of higher order in the curvature or other fields, as was schematically indicated in \redact.  Therefore we expect that these terms give contributions to the potential that involve higher powers of $1/R$.

Indeed, some leading corrections contributing to the radial dilaton potential for the flux compactifications of \GKP\ were discussed in \refs{\BBHL}.   For the case of a supersymmetry breaking vacuum, they find a four-dimensional potential that behaves as
\eqn\beckercor{V_K\propto {1\over R^{18}}}
at large $R$.  

\newsec{Problems for quintessence}

Given the apparent instability of a theory with extra dimensions and a positive cosmological constant, an obvious question is whether we could be presently experiencing this instability.  Specifically, if there is a non-trivial potential for the radial modulus, we might ask whether the Universe could be following a rolling solution in this potential, for example rolling to $R=\infty$ in  the potential of fig.~1.  Indeed, if this were consistent with observation, that would obviate the need for a local de Sitter minimum in the potential.

The key point in order to realize such a scenario is that the potential decreases slowly enough.  In particular, \refs{\HKS,\FKMP} have argued that this is difficult to achieve in string theory.  Let us reexamine this question for general radial dilaton potentials such as found in eq.~\redact.

This problem has been investigated by Ratra and Peebles in \refs{\RaPe}.  Specifically, they consider an action of the form
\eqn\PRact{ {M_4^2\over 16\pi}\int d^4x\sqrt{-g}\left[\cR_4 -{1\over 2} (\nabla \phi)^2 -{1\over 2} V(\phi)\right]\ .}
For an exponential potential,
\eqn\exppot{V=Ae^{-a\phi}\ ,}
they find ``tracker" solutions obeying the equation of state
\eqn\eqstate{p=w\rho}
(with constant $w$), and argue that more general solutions asymptote to these.  The necessary condition for an accelerating universe, $w<-1/3$, is obtained for $a<1$.  Working with the field $D$, with normalization given in eq.~\redact, and exponential potential
\eqn\Dpot{V(D)\propto e^{-bD}\ ,}
the corresponding condition for acceleration is
\eqn\qunitcond{b<\sqrt{d(d+2)}\ .}
In other words, the action for the matter content generating the potential must grow at least as fast as 
\eqn\actgrow{R^{2d-\sqrt{d(d+2)}}}
as $R\rightarrow\infty$.

Comparing to the results of section 2.2, we see that quintessence is difficult to achieve.  Eq.~\fluxpot\ shows that flux-generated potentials fall too fast at infinity.  Likewise, eq.~\branepot\ shows that for wrapped branes, the only case where acceleration can be achieved is when the brane wraps {\it all} of the dimensions, or in other words, when there is a $D$-dimensional cosmological constant present.\foot{This point was realized in collaboration with M. Lippert.}  Decompactification is in general not benign, and typically looks nothing  like a slowly rolling quintessential scenario.

\newsec{Discussion}

The analysis of section two gives a very general argument that, if we are living in a vacuum with non-zero cosmological constant, this vacuum is generically unstable.  There are several possible outcomes.  The most generic one is if the potential is positive with vanishing asymptotic behavior at infinite volume, and thus four-dimensional space ultimately has a decompactification instability.
Another possibility is that there is an AdS minimum in the radial dilaton potential; in this case the Universe may ultimately tunnel to that minimum and undergo gravitational collapse in a big crunch. 
 
A final possibility is if there are exactly supersymmetric zero-energy compactifications to four dimensions; these would then be degenerate with the decompactified vacuua.  Until recently many string theorists may have questioned this possibility, believing that some non-perturbative physics 
should lift these minima and yield a solution with supersymmetry broken at the TeV scale.  The puzzle was how to find such a vacuum with a zero cosmological constant, but recent developments have clearly lessened the motivation for finding vacuua with cosmological constants less than that presently observed!  If such vacua do exist as exact solutions of string theory, it becomes a dynamical question whether they are favored over decompactified space.  Since, at least in known cases, the radial dilaton potential is generated by fluxes and wrapped branes, it would be a challenge to find a mechanism to eliminate these and safely land in a supersymmetric vacuum.  For example, in the models presented in \KKLT, to eliminate the SUSY breaking one would have to annihilate the anti-D3 branes\KaVe; this process is subdominant to tunneling through the moduli behavior.  Moreover, if the gauge dynamics of our world is realized in a brane-world scenario by states on the anti-D3 branes, this would be even more catastrophic than implied by merely suddenly restoring supersymmetry!

An even more exotic possibility would be if there are other dS vacua, with higher or lower vacuum energy, in which case we might be in the midst of a cascade.  As in \refs{\BoPo}, this would provide a possible framework for application of the anthropic principle.\foot{ Susskind\Sussanth\ has recently given an extensive discussion of the possible role of the anthropic principle in such a context.}

It is also worth emphasizing that even in the case of a de Sitter vacuum that is realized by a compact theory that is not well-described in the semiclassical approximation, one expects the same instabilities.  Indeed, this will be true as long as any candidate vacuum is continuously connected through a path in the parameter space of the theory to the semiclassical configurations discussed in section two. In string theory it is generally expected, though not proven, that this is the case.  If this is true, the Universe should reach the unstable configurations through quantum tunneling and/or thermal excitation.  The only obvious way to avoid the generic instability is if there are non-geometric solutions of string theory that are totally isolated from geometrical solutions.

The problem of computing the decay time is  discussed in some detail in ref.~\KKLT, and \refs{\GKS} distills a rough general argument for its size.  The probability of a fluctuation in de Sitter space that takes one to a configuration on top of the potential barrier intervening between the metastable de Sitter vacuum and neighboring configurations is of order 
\eqn\expds{e^{\Delta S}\ ,}
where $\Delta S$ is the difference in entropy between the two configurations.  If we think of the configuration at the top of the barrier as a solution with energy $V_{top}$, then the difference in entropy is
\eqn\entdiff{\Delta S = S_{top} - S_{dS} \propto {1\over V_{top}} -{1\over V_0} \ . }
This gives a time scale
\eqn\decayt{T_{decay} \sim e^{S_{dS} - S_{top}}\ .}
Ref.~\GKS\ observes that this time scale is always shorter than the recurrence time.  However, given that the recurrence time is long,
\eqn\recurtime{T_{recur}\sim \exp\left\{ 10^{120}\right\}\ ,}
and the potential barrier is expected to be substantial,
this is very unlikely to create a sense of urgency.

The generic instability of de Sitter solutions in theories with extra dimensions does, therefore, render the physics of Poincar\'e recurrences irrelevant.  Indeed, this generic instability appears  to offer an elegant resolution of  the various theoretical conundrums that an eternal de Sitter space seemingly implies.

Many interesting problems remain.  Top among them is finding an explanation for why our metastable de Sitter minimum gives such a small vacuum energy.  For this one may even have to seek an anthropic explanation, as in \BoPo.  An important first step in such an explanation would be a detailed theory predicting a large number of possible minima with widely varying vacuum energies.  This raises the interesting possibility that the stage of inflation that preceded our present cosmological epoch was merely another step in a cascade between de Sitter minima.  Whether or not this is the correct explanation of the smallness of the present vacuum energy, another challenge is to find such compactifications with viable phenomenology.  In string theory, we have probably just scratched the surface in exploring the space of solutions with branes and fluxes.

By now we have come to appreciate that extra dimensions have many virtues.  But, when combined with present cosmological data indicating a positive vacuum energy, we learn that they also seemingly doom our four-dimensional Universe to a catastrophic instability.  The instability generically leads to growth of the extra dimensions of space, although it can result in a big crunch, or, possibly,  subsequent cosmological expansion in a supersymmetric four-dimensional world. On the positive side, the decay can result in a state that does not suffer the ultimate fate of infinite dilution and thermalization implied by de Sitter space.  We can seek solace both in the relatively long life of our present four-dimensional Universe, and in the prospect that its decay produces a state capable of sustaining interesting structures, perhaps even life, albeit of a character very different from our own.

\bigskip\bigskip\centerline{{\bf Acknowledgments}}\nobreak

I'd like to thank O. Aharony, O. DeWolfe, G. Horowitz, C. Johnson, S. Kachru, 
J. Polchinski, M. Srednicki, and particularly M. Lippert and L. Susskind for very valuable conversations.
This work was
supported in part by the Department of Energy under
Contract DE-FG-03-91ER40618, and by the George P. \& Cynthia W. Mitchell Institute for 
Fundamental Physics, Texas A\&M University.

\listrefs
\end